\documentclass[showpacs,amsmath,amssymb,twocolumn,prx,longbibliography,superscriptaddress,preprintnumbers,nofootinbib,superscriptaddress]{revtex4-2}
\makeatletter
\def\fnum@figure{FIG.~\thefigure}
\makeatother
\usepackage{amssymb}
\usepackage[dvips]{graphicx}
\usepackage{enumerate}
\usepackage{epsfig}
\usepackage{subfigure}
\usepackage{xcolor}
\usepackage[T1]{fontenc}
\usepackage{fullpage}
\usepackage{amsthm,amsfonts,amssymb,amscd,mathrsfs,xspace,framed}
\usepackage{booktabs}
\usepackage{siunitx}
\usepackage{threeparttable}
\usepackage{mathrsfs,amsmath}
\usepackage[linguistics]{forest}
\newcommand{\okina}{\textquotesingle}
\usepackage{color}
\usepackage{enumitem}
\usepackage{setspace}
\usepackage{url}
\usepackage{eqparbox} 
\usepackage{wrapfig}
\usepackage{array,multirow}
\usepackage{enumitem}
\usepackage{bm}
\usepackage{comment}
\usepackage{txfonts}
\renewcommand{\arraystretch}{1.3}
\usepackage{upgreek}
\newtheoremstyle{nonitalic}
  {3pt}   
  {3pt}   
  {\normalfont}  
  {}      
  {\bfseries}  
  {.}     
  { }     
  {}      

\theoremstyle{nonitalic}

\usepackage{braket}
\usepackage{mathtools}

\usepackage[all]{xy}

\usepackage[hyperfootnotes=false]{hyperref}
\usepackage[acronym]{glossaries}
\usepackage[english]{babel}
\usepackage{url}
\usepackage{bm}
\usepackage{hyperref}
\definecolor{darkblue}{rgb}{0,0,0.5}
\hypersetup{
colorlinks=true,
linkcolor=black,
filecolor=blue,
citecolor=darkblue,  
urlcolor=black,
}
\usepackage{gensymb}

\usepackage{xcolor}

\begin{document}

\title{Microring Perceptron Sensing for Low-Power Radio-Frequency Detection with\\Quantum-Compatible Photonic Preprocessing}

\author{Bo-Han Wu$^\S$}
\email{bohanwu@hawaii.edu}
\affiliation{Research Laboratory of Electronics, Massachusetts Institute of Technology, Cambridge, Massachusetts 02139, USA}
\affiliation{Electrical and Computer Engineering, University of Hawai\okina i at M\={a}noa, Honolulu, Hawai\okina i 96822, USA}

\author{Shi-Yuan Ma$^\S$}
\email{mashiyua@mit.edu}
\affiliation{Research Laboratory of Electronics, Massachusetts Institute of Technology, Cambridge, Massachusetts 02139, USA}

\author{Mahmoud Jalali Mehrabad}

\author{Mingran Jia}

\author{Sri Krishna Vadlamani}
\affiliation{Research Laboratory of Electronics, Massachusetts Institute of Technology, Cambridge, Massachusetts 02139, USA}

\author{Hyeongrak Choi}
\affiliation{Research Laboratory of Electronics, Massachusetts Institute of Technology, Cambridge, Massachusetts 02139, USA}
\affiliation{Department of Electrical and Computer Engineering, Stony Brook University, New York 11794, USA}

\author{Dirk Englund}
\affiliation{Research Laboratory of Electronics, Massachusetts Institute of Technology, Cambridge, Massachusetts 02139, USA}

\date{\today}

\begin{abstract}
Radio-frequency (RF) sensing underpins applications ranging from radar and wireless communication to biomedical and quantum measurement, where detection sensitivity at low signal-to-noise ratio (SNR) directly limits the achievable range, resolution, and information capacity. Machine learning has been widely applied to enhance sensing performance, but predominantly as post-detection analysis of already-acquired data. At low SNR, however, the information missing from acquired data cannot be recovered downstream, making pre-detection processing essential. Here we introduce microring perceptron (MiRP) sensing, a framework that exploits RF-photonic transduction in a microring resonator, where a programmable optical pump performs three-wave mixing to implement a learned mapping from the incoming RF signal to optical feature signals prior to detection. The transducer front end and digital neural-network back end are jointly optimized within a single end-to-end training pipeline, allowing the sensing system to learn an encoding that preserves task-relevant information through detection. Across benchmark tasks, MiRP sensing achieves substantially higher task performance than conventional processing at low input RF power levels where detection noise dominates. The algorithmic gain reported here offers an orthogonal axis of improvement that composes with, and amplifies the impact of, existing and future advances along complementary dimensions, including hardware efficiency and quantum-enhanced optical sensing.

\end{abstract}

\maketitle
\def\thefootnote{\S}\footnotetext{These authors contributed equally to this work}\def\thefootnote{\arabic{footnote}}

\section{Introduction}

Sensing across the electromagnetic spectrum, from optical and infrared to terahertz, microwave, and radio frequencies, underpins how we observe, communicate with, and control the physical world. Within this spectrum, radio-frequency (RF) sensing supports a particularly broad range of modern technologies, including wireless communication~\cite{alamouti1998simple}, radar~\cite{skolnik1980introduction}, remote sensing~\cite{curlander1991synthetic}, and the readout of quantum devices~\cite{dane2025,warner2023coherent}. A common thread across these applications is the need to recover weak RF signals from a noisy background: long-range radar returns, faint communication signals at the edge of coverage, and dispersive readout of superconducting qubits all operate at low signal-to-noise ratio (SNR), where receiver sensitivity directly limits performance. Conventional electronic receivers address this regime with superheterodyne architectures that amplify and down-convert the waveform before digitization~\cite{Pozar05}, but thermal noise introduced at the receiver fundamentally bounds the recovery of signals near or below the noise floor~\cite{van2001detection,van2004detection}.

\begin{figure*}
    \centering
	{\centering\includegraphics[width=0.95\linewidth]{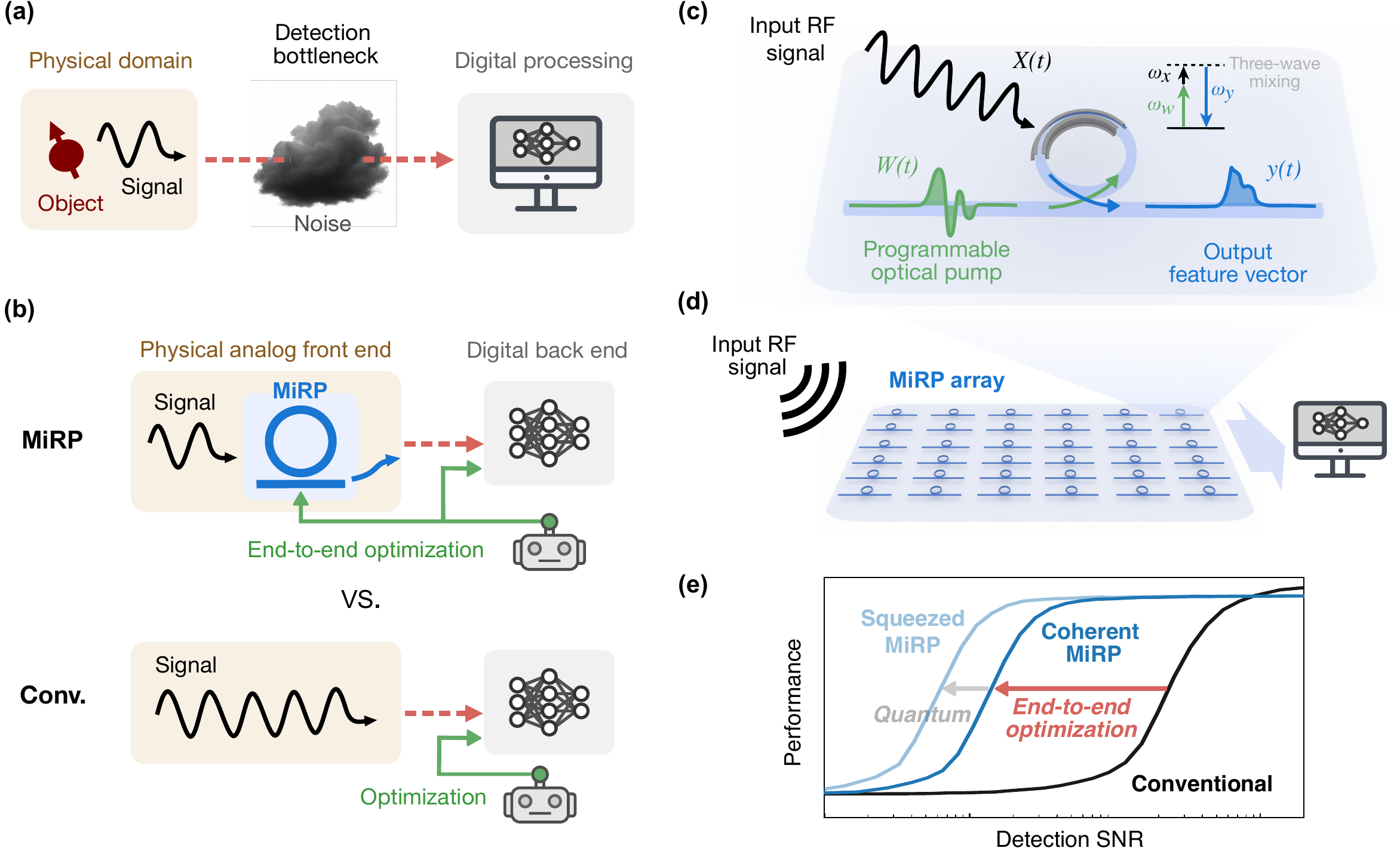}}
	\caption{Detection bottleneck in RF Sensing and the quantum-enhanced MiRP architecture. \textbf{(a)} General sensing scheme and the detection bottleneck. The information from the signal is affected by the background and sampling noise at the detection stage. \textbf{(b)} MiRP sensing pipeline: the signal is first processed by a MiRP physical analog front end and optically encoded prior to digital processing; conventional sensing pipeline (Conv.): the signal is directly measured and digitized for digital processing. \textbf{(c)} Physical layout of MiRP unit. A programmable coherent optical field $W(t)$ is injected into each MiR to encode features of the input waveform $X(t)$, producing an optical output $y(t)$. \textbf{(d)} MiRP array layout. \textbf{(e)} Performance advantage of coherent and squeezed MiRP relative to the conventional approach as a function of detection SNR.}
	\label{fig:Architecture}
\end{figure*}

Recently, machine learning has become a widely adopted tool for RF sensing~\cite{santra2025machine}, enabling data-driven inference from complex RF measurements across tasks such as modulation classification~\cite{o2016convolutional,mendis2016deep} and parameter estimation~\cite{ye2017power,kumar2024analysis,sarkar2024radyololet}.
As in most AI-for-science applications, these methods operate on the samples produced after detection, and the digital model is the locus of optimization. The performance of any such method is therefore bounded by the accessible data that survives detection. When the SNR at detection is low, noise irreversibly corrupts task-relevant features of the waveform, resulting in the \emph{detection bottleneck}~\cite{ma2026machine}, where a significant fraction of the signal's information is lost (see Fig.~\ref{fig:Architecture}\textbf{(a)}). By the data processing inequality, no subsequent digital processing can restore what has already been lost~\cite{beaudry2011intuitive}. Preserving this information, therefore, requires intervention in the physical front end of the sensing pipeline, before detection.

Physical-domain preprocessing of RF signals before detection is itself an active area, with implementations spanning programmable metasurfaces and reconfigurable apertures~\cite{hunt2013metamaterial,jiang2024simultaneously,del2020learned,li2020intelligent}, microwave-photonic feature extractors~\cite{Xu2024,zhang2023broadband,zhang2024system,sinigaglio2026photonic}, and emerging platforms such as spintronic and neuromorphic RF hardware~\cite{ross2023multilayer,gao2023programmable,zheng2023neuroradar,senanian2024microwave}. The performance of any such preprocessing front end can be further improved by jointly optimizing it with the digital back end for a given task, so that the front end shapes the signal into a representation matched to the downstream model rather than being designed in isolation. End-to-end optimization of this kind has been developed extensively in computational imaging and optical sensing~\cite{sitzmann2018end,metzler2020deep,tseng2021neural,deb2022fouriernets,pinkard2024information,ma2026machine}, and has also been brought to RF sensing using trainable metasurface transceivers~\cite{del2020learned,li2020intelligent}. 
On the RF-photonic side, however, the performance advantage of treating the physical front end as a learnable component within the same training architecture as the digital back end has not been clearly established, particularly in the low-power regime where detection noise dominates.

Here we propose the microring perceptron (MiRP) sensing framework (see Fig.~\ref{fig:Architecture}\textbf{(b)}), which leverages a microring (MiR) transducer driven by a programmable optical pump to realize a learnable RF-photonic receiver. The optical pump performs analog preprocessing of the incoming RF waveform via three-wave mixing prior to photodetection, and is jointly trained with the digital back-end neural network in an end-to-end fashion (Fig.~\ref{fig:Architecture}\textbf{(c,d)}). This joint optimization is the central contribution of MiRP: at a fixed detection SNR, it yields higher task performance than conventional RF receivers that digitize the raw waveform, and the advantage applies independently of the specific RF-to-optical transduction technology used to realize the front end (Fig.~\ref{fig:Architecture}\textbf{(e)}). Beyond this core algorithmic advance, the RF-photonic implementation carries two additional considerations. First, although RF-to-optical transduction is often regarded as inefficient, integrated MiR-based receivers achieve power efficiencies comparable to those of electronic receivers of similar footprint (Appendix~\ref{sec:Transduction_eff}), so the algorithmic gain comes at no hardware cost. 
Second, because measurement is performed in the optical domain, MiRP composes naturally with established optical-sensing techniques such as quantum  enhancement with squeezed light~\cite{Aasi2013Enhanced,Yu2020Quantum,yap2020broadband,mcculler2020frequency} (light blue curve in Fig.~\ref{fig:Architecture}\textbf{(e)}), without modification of the core architecture.

\section{The MiRP Sensing Framework}
\label{sec:framework}

\subsection{Programmable RF-to-optics transduction}
\label{sec:transduction}
The unit of MiRP is a high-$\chi^{(2)}$ MiR transducer whose refractive index is modulated by the input RF signal (see Fig.~\ref{fig:Architecture}\textbf{(c)}; full design details in Appendix~\ref{sec:EO_transducer}). The input field $X(t)\,e^{-i\omega_x t}$, with envelope $X(t)$ and carrier frequency $\omega_x$, drives a voltage across an on-chip LC resonator capacitor, coupling to the MiR via the transduction Hamiltonian
\begin{equation}
    \hat{\mathcal{E}}(t)=\frac{i\hbar \,\mu}{2} \,\left(\hat{x}^{\dagger}(t)+\hat{x}(t)\right)\,X(t),
    \label{eq:Hamiltonian_encoding}
\end{equation}
where $\hbar$ is the reduced Planck's constant, $\mu\in\mathbb{R}$ is the transduction coupling rate, and $\hat{x}(t)$ is the annihilation operator of the ring RF mode. Simultaneously, a time-binned optical pump $W(t)\,e^{i\omega_w t}$, where $W(t)$ encodes a sequence of discrete temporal modes, is injected into the MiR. The $\chi^{(2)}$ nonlinearity mediates three-wave mixing (TWM) between these two fields with Hamiltonian
\begin{equation}
    \hat{\mathcal{H}}(t) = \frac{i\hbar\,g}{2} \, W(t) \,\bigl(\hat{x}^{\dagger}(t)\,\hat{a}(t) + \hat{x}(t)\,\hat{a}^{\dagger}(t)\bigr),
    \label{eq:Hamiltonian_EO}
\end{equation}
where $g\in\mathbb{R}$ is the electro-optic coupling rate and $\hat{a}(t)$ is the output annihilation operator at frequency $\omega_y=\omega_w+\omega_x$. 

Solving the Heisenberg--Langevin equations yields the output envelope  (see Appendix~\ref{sec:analytical} for details) 
\begin{equation}
    y(t) = \left\langle\left(\hat{a}(t)+\hat{a}^\dagger(t)\right)\right\rangle \propto W(t)\,X(t).
    \label{eq:output_envelope}
\end{equation}
Each pump time bin samples the RF envelope and weights it by a learned amplitude, producing a sequence of encoded values that form a feature vector. The output of each channel, $y(t)$, is ultimately measured via balanced homodyne detection.

\subsection{MiRP array}
To capture the full temporal structure of $X(t)$, $L$ MiRP units are integrated (see Fig.~\ref{fig:Architecture}\textbf{(d)}), where each MiRP unit is driven by an optical pump $W_l(t)$ corresponding to the $l$-th learned waveform, with $l=0,\cdots,L$. Each $W_l(t)$ acts as a temporal kernel, implementing the multiplicative encoding of Eq.~\eqref{eq:output_envelope} to produce an output feature $y_l(t)\propto W_l(t)\,X(t)$. Collectively, MiRP carries out $L$ learned kernels that extract complementary features of the input signal. Because the RF wavelength ($\sim1$~cm) far exceeds the single MiR dimension ($\sim0.1$~mm), the incident field uniformly illuminates all elements without active fan-out, enabling simultaneous access to the same input while generating parallel feature outputs $\{y_l(t)\}_{l=1}^L$ through distinct optical encodings.

\subsection{Optimization across the physical--digital interface}
\label{sec:optimization}
The mapping in the physical front end, $X(t)\rightarrow y_l(t)$, is determined by the pump waveforms $W_l(t)$. We treat this mapping as a trainable layer as in the entire physical--digital optimization. This enables end-to-end training of the pump waveforms, allowing the analog encoding to be directly optimized for the downstream task.

The trainable physical front end is followed by convolutional and fully connected digital layers. Training is performed on clean (noise-free) input signals; the detection noise model is applied only during testing, so the learned pump waveforms are not tuned to a particular SNR but instead feature encodings that are inherently robust across a wide range of noise conditions.

To isolate the effect of end-to-end optimization, we compare MiRP against a baseline in which the pump is replaced by a continuous-wave (CW) field matched in average optical power, corresponding to the conventional operating mode for a MiR transducer. At a given detection SNR, any performance gap between MiRP and this baseline is therefore purely algorithmic: it reflects the benefit of learned analog preprocessing prior to digitization. This baseline constitutes the conventional RF-photonic (RFP) sensing configuration.

\subsection{From detection SNR to the input RF power}
\label{sec:snr_to_power}
While detection SNR is the natural variable for evaluating algorithmic contribution, practical sensing systems are typically characterized by input power level. We bridge these perspectives via the power-to-SNR coefficient $\alpha$, determining the detection SNR by
\begin{equation}
    \text{SNR} = \alpha\,P_\text{RF},
    \label{eq:SNR_power}
\end{equation}
which absorbs the relevant receiver physics into a single scalar and enables direct comparison across sensing platforms (see derivations in Appendix~\ref{sec:Transduction_eff}).

We next examine two complementary mechanisms, hardware power efficiency and quantum compatibility.
 
\subsubsection{Hardware power efficiency}
\begin{figure*}
    \centering
{\centering\includegraphics[width=0.95\linewidth]{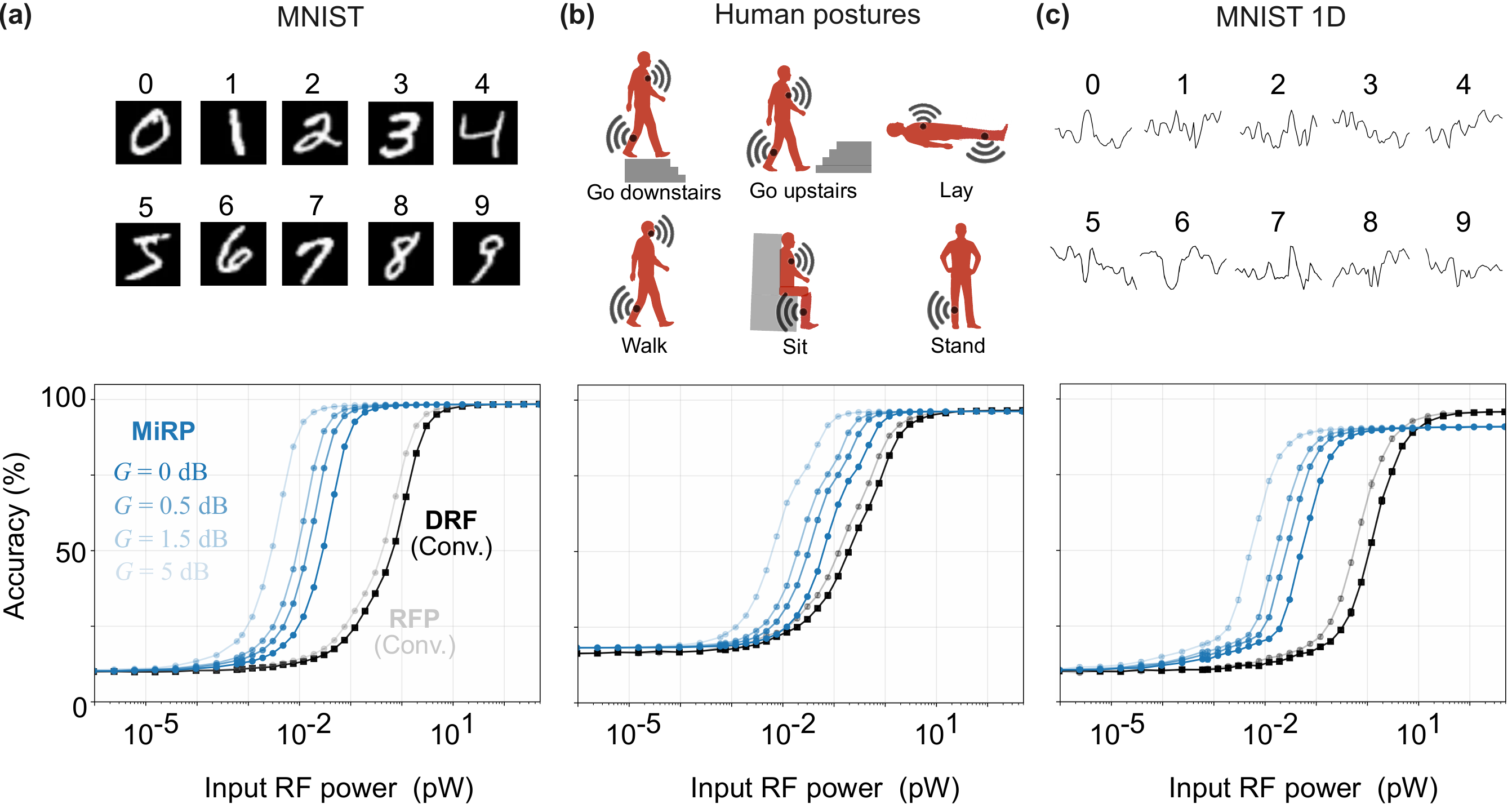}}
	\caption{Classification performance across three datasets. Top: representative samples from each dataset. Bottom: classification accuracy versus input RF power. \textbf{(a)} MNIST~\cite{lecun2002gradient}. \textbf{(b)} Human posture~\cite{anguita2013har}.
\textbf{(c)} MNIST-1D~\cite{greydanus2024scaling}. 
Blue curves show MiRP at four squeezing levels $G=0,0.5,1.5,5$~dB (darkest to lightest blue), where $G$ is the squeezing-induced SNR gain in dB. Black and gray curves show the direct-RF (DRF) and RF-photonic (RFP) conventional baselines, which share the same trained digital model and differ only in their hardware efficiency. MiRP is optimized end-to-end across the physical front end and digital back end.}
	\label{fig:performance}
\end{figure*}

Any realistic receiver has a finite power efficiency from the input to the detection stage, determined by platform-specific factors such as noise figure and bandwidth in electronic receivers or transduction efficiency $\beta$ in RF-photonic systems. A given detection-SNR performance curve therefore maps to a different power axis for each platform.

This mapping has two baselines: (1) RFP sensing, uses the same receiver as MiRP and therefore the same coefficient $\alpha_\text{M}$ (see Eq.~\ref{eq:SNR_def}). (2) DRF sensing, employs a superheterodyne receiver characterized by $\alpha_\text{D}$ (see Eq.~\ref{eq:SNR_LM}), providing the standard RF sensing reference. The horizontal separation between the RFP and DRF conventional curves on the power axis reflects only the hardware ratio $\alpha_\text{M}/\alpha_\text{D}$, which we evaluate against state-of-the-art integrated receivers in Section~\ref{sec:results}. Critically, even on a platform where $\alpha_\text{M} < \alpha_\text{D}$, MiRP's learned analog preprocessing remains an independent, orthogonal dimension of performance improvement.

\subsubsection{Quantum compatibility}
\label{sec:quantum_compat}
Beyond classical power efficiency, the RF-photonic MiRP architecture is intrinsically compatible with quantum-enhanced sensing~\cite{taylor2013biological,Aasi2013Enhanced,Yu2020Quantum,tse2019quantum}. Injecting squeezed light in place of a coherent pump implements a phase-sensitive amplifier (PSA) with gain $G=\cosh^2 r$, yielding a multiplicative SNR enhancement of $G$ with the same optical power (Appendix~\ref{sec:analytical}). Squeezing-based enhancement is a well-established technique across diverse sensing platforms~\cite{dutt2015chip,yang2021squeezed,zhao2020near,liu2025wafer,ren2026quantum}. A common concern with quantum-enhanced sensing is that fragile nonclassical states degrade rapidly under loss, limiting practical utility~\cite{Zhang2015PRL}. MiRP avoids this regime: the squeezed light is confined within the sensing system rather than transmitted as a lossy probe, making the enhancement robust in practice (e.g., Ref.~\cite{Aasi2013Enhanced}).

Hardware power efficiency and quantum noise suppression are independent, multiplicative factors that together determine the mapping from detection SNR to input RF power. MiRP's learned analog preprocessing (Section~\ref{sec:optimization}) operates along a third orthogonal axis and composes freely with both, so gains from improved transduction hardware and from quantum resources carry over directly to the MiRP-enhanced system.

\section{Performance Evaluation}
\label{sec:results}
We evaluate MiRP on three classification tasks, benchmarked against two conventional baselines that share the same digital back end: the RF-photonic receiver without learned preprocessing (i.e., RFP, identical hardware to MiRP driven by a CW coherent pump) and a digital superheterodyne receiver (i.e., DRF). In all cases, task information is encoded in the temporal envelope of the input RF field $X(t)$, performance is reported as classification accuracy, and the input RF power $P_\text{RF}$ spans $1$~aW to $0.5$~nW. Across every task, \textit{coherent} MiRP (i.e., $G=0$~dB) retains near-ceiling accuracy at input powers where both conventional baselines have already collapsed to near-chance performance (see Fig.~\ref{fig:performance}). Incorporating squeezed light yields further improvements, with \textit{squeezed} MiRP ($G>0$~dB) achieving higher accuracy under the same conditions (see Eq.~\eqref{eq:Gain_effect}). 

\subsection{Evaluation setup}
\label{sec:config}
\subsubsection{Hardware calibration of $\alpha$} 
As developed in Section~\ref{sec:framework}, training operates purely in detection-SNR space, and each hardware platform enters through its power-to-SNR coefficient $\alpha$ via $\text{SNR} = \alpha\, P_\text{RF}$. For the results reported below we adopt representative values $\alpha_\text{M} = 6.2~\text{pW}^{-1}$ for the RF-photonic receiver (computed from the transduction efficiency $\beta = 1.18$\,\% of Ref.~\cite{warner2023coherent} in Tab.~\ref{tab:efficiency}) and $\alpha_\text{D} = 4.1~\text{pW}^{-1}$ for the DRF baseline (highest reported value for integrated superheterodyne receivers~\cite{zijlma2024} in Tab.~\ref{tab:superhet_sota}). Both correspond to state-of-the-art, fully integrated receivers of comparable chip-scale footprint, for which the two platforms are broadly competitive (physical parameters in Tab.~\ref{tab:parameters}; derivation in Appendix~\ref{sec:Transduction_eff}). Because MiRP and RFP use the same RF-to-optics transduction hardware, they share $\alpha_\text{M}$: any performance gap between them isolates the algorithmic contribution of the learned preprocessing $W(t)$.

\subsubsection{Digital back-end neural network} 
To ensure that any MiRP advantage arises from the inclusion of the physical layer during optimization rather than from differences in digital capacity, all three schemes share the same digital-layer structure, consisting of two convolutional layers with max-pooling followed by three fully connected layers mapping to class predictions, while the layer dimensions are independently optimized for each sensing pipeline and task. MiRP additionally includes the physical front end prior to detection. The RFP and DRF curves correspond to the same digital model evaluated through different values of $\alpha$ and are not independently trained systems.

\subsection{Classification results}
We evaluate MiRP on three classification tasks: MNIST~\cite{lecun2002gradient}, human posture detection~\cite{anguita2013har}, and MNIST-1D~\cite{greydanus2024scaling} (see Appendix~\ref{sec:tasks}). We first verify at high RF power that all three schemes share similar digital capacity, then turn to the low-power regime where MiRP's learned preprocessing is intended to matter.

\begin{figure*}
    \centering
{\centering\includegraphics[width=0.9\linewidth]{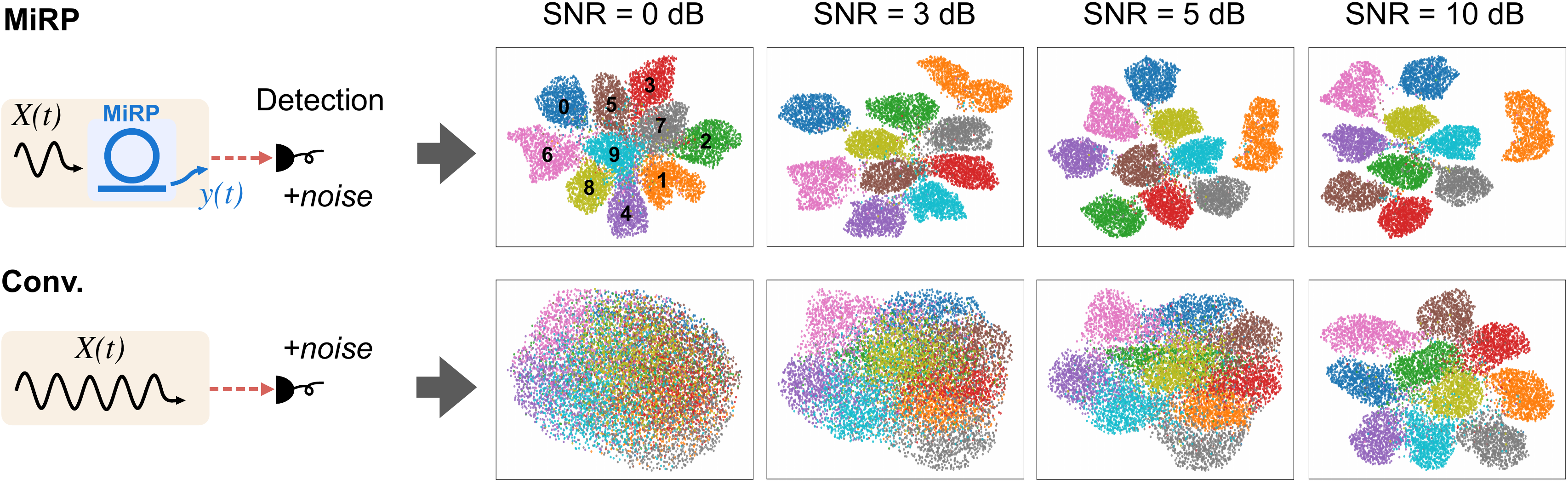}}
	\caption{Two-dimensional feature-space visualization using UMAP~\cite{mcinnes2018umap} embeddings of detected MNIST waveforms at detection SNRs of $0$, $3$, $5$, and $10$~dB. Top row: output feature vectors of the MiRP front end, $y(t)$. Bottom row: directly detected input RF signal, $X(t)$. For each row, the singal vectors ($y(t)$ or $X(t)$) are combined with additive noise at the indicated SNR, and the resulting noisy vectors are then visualized by UMAP. Colors denote MNIST classes, with class indices labeled in the $\text{SNR}=0$~dB  MiRP panel. The MiRP-transformed features remain class-separable down to 0~dB, whereas the input RF signal shows class structure only at the highest SNR shown.}
	\label{fig:diagnosis}
\end{figure*}

\subsubsection{High RF power: matched digital capacity}
In the high-power regime where detection noise is negligible, all three schemes converge to the widely reported architectural ceiling on each task, reaching $\sim 98$\,\% accuracy on MNIST (see Fig.~\ref{fig:performance}(a)). This confirms that MiRP's analog preprocessing does not discard task-relevant information: the transformation it applies in the physical domain leaves the digital back end with enough to reach the same ceiling as the conventional schemes. On a more challenging task, MNIST-1D, where a digital MLP can only have $\sim60\%$ accuracy~\cite{greydanus2024scaling}, the conventional schemes slightly exceed MiRP for $P_\text{RF}\gtrsim 0.1$~nW (see Fig.~\ref{fig:performance}(c)). This gap reflects a structural limitation of the physical layer: MiRP implements a transformation that, unlike the convolutional layers of the digital back end, does not exploit the spatial correlations to which MNIST-1D is particularly well matched. This small high-power penalty is not at odds with the MiRP framework: MiRP does not claim greater expressivity than a purely digital back end in the high-power regime. MiRP's intended regime of advantage is the low-power, low-SNR regime, where information loss at detection creates a bottleneck (see Fig.~\ref{fig:Architecture}\textbf{(a)}) that no digital processing can undo. Taken together, these observations rule out excess digital capacity as the source of MiRP's low-SNR advantage.

\subsubsection{Low RF power: pushing back the detection bottleneck}
\label{sec:classification}
As shown in Fig.~\ref{fig:performance}, MiRP without squeezing ($G=0$) maintains near-ceiling accuracy at input powers where both conventional baselines have already degraded. On MNIST at $P_\text{RF} \sim 1$~pW, MiRP attains the high-SNR ceiling while DRF and RFP have already dropped significantly. The same qualitative pattern holds on human posture detection and MNIST-1D: the power threshold at which each scheme reaches $\sim 90\%$ of its ceiling accuracy is shifted by roughly two orders of magnitude between MiRP and DRF. 

This total gap decomposes cleanly along the two axes of the framework. The RFP–DRF separation on the power axis is fixed by the hardware ratio $\alpha_\text{M}/\alpha_\text{D}$ alone, since both schemes apply the same digital model through different transduction coefficients. The MiRP–RFP separation, in contrast, occurs at fixed hardware and therefore isolates the contribution of the learned preprocessing: the pump waveform $W(t)$ is optimized end-to-end to concentrate task-relevant structure into modes that remain resolvable above the detection noise floor, so discriminative information survives the detection bottleneck that irreversibly removes it for conventional schemes. Squeezed MiRP extends this advantage along an axis independent of the learned transformation. Injecting squeezed vacuum with gain $G$ scales the effective detection SNR (see Eq.~\eqref{eq:Gain_effect}), translating each MiRP accuracy curve leftward on the power axis by $G$. Because squeezing acts on the optical pump rather than on the learned transformation, its gain composes freely with the algorithmic advantage rather than substituting for it.

\subsection{Feature representation at the detection boundary}
\label{sec:feature}

To get better intuition where the low-SNR advantage of MiRP originates, we visualize the feature representations passed to the digital back end at a range of detection SNR levels. Fig.~\ref{fig:diagnosis} shows two-dimensional UMAP~\cite{mcinnes2018umap} of the original MNIST input vectors $X(t)$ and the MiRP-transformed feature vectors $y(t)$ under varying detection noise. At high detection SNR (e.g., $10$~dB), both representations produce ten well-separated clusters, indicating that class-discriminative structure is preserved in each case. As the detection SNR decreases to $5$, $3$, and $0$~dB, the clusters formed from sampled $X(t)$ progressively degrade into a noisier and more diffuse distribution, while those formed from $y(t)$ remain clearly separated across the same range. This contrast shows that the learned MiRP transformation concentrates class-discriminative information into modes that remain above the detection bottleneck, so that the digital back end receives a representation in which the classes remain separable after detection noise. The resulting gain is therefore not merely improved noise robustness but a restructuring of how information is organized across the analog-to-digital boundary.
Fig.~\ref{fig:diagnosis} thus makes the algorithmic claim concrete: at matched detection SNR, MiRP preserves class structure that conventional detection has already lost, translating directly into the accuracy gap of Fig.~\ref{fig:performance}.

\section{Discussion}
\label{sec:discussion}
In this paper, MiRP demonstrates that jointly optimizing a programmable optical front end with a digital back-end neural network can substantially mitigate the detection bottleneck in low-power RF sensing. At low detection SNR, any processing pipeline that operates exclusively on the detected data suffers severe degradation. MiRP instead intervenes at the physical front end by shaping the analog feature space through pump waveforms that are jointly optimized with the digital back-end neural network, encoding task-relevant structure into the optical field before information is lost to detection noise. 
Across the benchmark tasks evaluated here, this yields consistent and substantial performance gains over conventional sensing across a range of low-power, detection-noise-limited tasks, with the advantage most pronounced at the lowest input power levels where the detection bottleneck is most severe. More broadly, MiRP exemplifies a design principle applicable beyond RF sensing: the effective place to embed machine learning is not in the digital back end alone, but at the physical interface where information is most at risk.


The algorithmic gains demonstrated here do not exist in isolation. MiRP is designed to compose naturally with three active and independent lines of development in the broader sensing community. The first is hardware efficiency: advances in RF-to-optical transduction directly raise the coefficient $\alpha$ and shift MiRP's performance curve to lower input powers, without touching the learned preprocessing. The second is quantum resources: techniques such as squeezing~\cite{Aasi2013Enhanced,Yu2020Quantum,yap2020broadband,mcculler2020frequency} and entanglement~\cite{Xia20,Zhuang22,Wu_2023,PhysRevA.97.032329,brady2023entanglement,hariri2024entangled}, developed extensively in the quantum sensing community, improve the effective SNR at a given detection power and integrate seamlessly into the homodyne readout already employed by MiRP. The third is integrated photonics scalability for the $\chi^{(2)}$ material: the rapid maturation of thin-film lithium-niobate~\cite{zhang2017monolithic,zhang2019broadband,bruch2019chip,yu2022integrated} and -tantalate platforms~\cite{wang2024lithium,zhang2025ultrabroadband} provides a direct pathway to large MiRP arrays, expanding the dimensionality of the extractable feature space as fabrication capabilities grow.

Advances in receiver hardware directly raise $\alpha$ by power efficiency, shifting MiRP's performance to lower input powers independently of the learned preprocessing.
Among current implementations, the transducer reported in~\cite{warner2023coherent} (Tab.~\ref{tab:efficiency}) provides the most suitable efficiency-bandwidth trade-off for MiRP, with $\beta = 1.18$~\% at a $30$~MHz bandwidth compatible with the signal bandwidths considered here (Fig.~\ref{fig:performance}). 
More broadly, the current transducer landscape presents a characteristic efficiency-bandwidth trade-off: electro-optomechanical devices reach higher efficiencies at kilohertz bandwidths~\cite{brubaker2022optomechanical,higginbotham2018harnessing}, while other electro-optic platforms offer broader bandwidths at lower efficiencies~\cite{sahu2022quantum}. 
This trade-off provides flexibility across different sensing requirements, and any progress in transducer performance within this active field will directly translate to further power-axis gains for MiRP. 

While advances in receiver hardware raise $\alpha_\text{M}$, practical constraints in integrated photonic platforms impose a ceiling on usable optical power, limiting how far hardware efficiency alone can be pushed. Quantum resources offer a fundamentally different lever: techniques such as optical squeezing~\cite{Aasi2013Enhanced,Yu2020Quantum,yap2020broadband,mcculler2020frequency} and entanglement~\cite{Xia20,Zhuang22,Wu_2023,PhysRevA.97.032329,brady2023entanglement,hariri2024entangled} can further improve the effective SNR at a fixed detection power without raising the optical field strength, as demonstrated in our results (Fig.~\ref{fig:performance}). 
Because the RF signal is transduced onto an optical carrier and the nonclassical states remain confined within the sensing apparatus rather than transmitted as lossy probes to the target, MiRP avoids the practical concern about loss-resilience in quantum-enhanced optical sensing, making the enhancement robust in practice. 

The integrated photonic platform underlying MiRP provides a low-loss, stable on-chip environment, and its rapid development opens a direct path to further scaling of sensing capabilities. State-of-the-art integrated RF-photonic transducers already achieve power-to-SNR coefficients competitive with fully integrated superheterodyne receivers of comparable footprint (Appendix~\ref{sec:Transduction_eff}), establishing the platform as practically viable for RF sensing today. Building on this foundation, each MiRP unit in the array implements a distinct, optimized processing kernel, so the array collectively spans a high-dimensional, task-specific feature space that a single transducer channel cannot access. The compact footprint of integrated photonic devices makes large arrays practically realizable, allowing the expressive power of the front end to scale with array size without a proportional increase in system complexity.

\begin{acknowledgments}
The authors thank Tingjun Chen, Zhihui Gao, Sheng-Xiang Lin, Haowei Shi, and Kfir Sulimany for fruitful discussions. B.-H.W. acknowledges support from Honda Research Institute USA, Inc., S.-Y.M. acknowledges support from Honda Research Institute USA, Inc. and the DARPA INSPIRED program. S.K.V. acknowledges support from TSMC. H.C. acknowledges support from Cisco Quantum Lab, MITRE, and Claude Shannon Fellowship. D.E. acknowledges funding from Honda Research Institute USA, Inc., the DARPA INSPIRED and the DARPA QuANET programs.
\end{acknowledgments}

\appendix
\begin{figure}
    \centering	{\centering\includegraphics[width=1\linewidth]{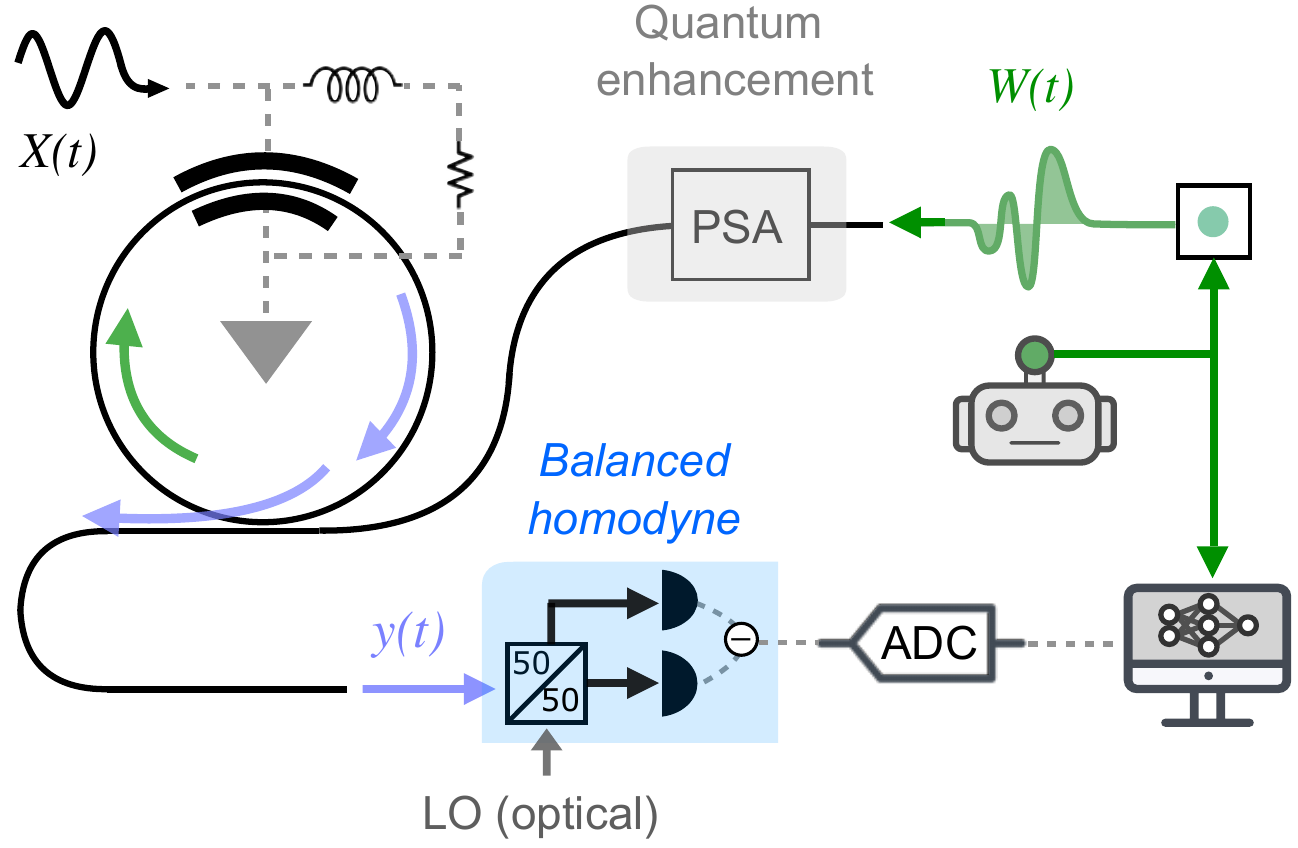}}
	\caption{Layouts of MiRP unit. The input RF $X(t)$ is transduced as a voltage on an LC resonator that modulates a high-$\chi^{(2)}$ MiR with a radiation resistance $R_\text{RF}$. The MiR is driven by a programmable optical pump $W(t)$, outputting optical field $y(t)$ (via TWM), followed by a balanced homodyne detector, ADC, and digital back end. The waveform $W(t)$ and digital back end jointly optimized with machine learning. PSA: phase sensitive amplifier. LO: local oscillator. ADC: analog-to-digital converter.}
	\label{fig:scheme}
\end{figure}

\section{Simulation parameters summary}

 MiRP's physical parameters are detailed in Tab.~\ref{tab:parameters}.

\renewcommand{\arraystretch}{1.3}
\begin{table}[h]
    \begin{tabular}{c @{\hspace{1.3em}} c @{\hspace{1.9em}} l}
        \multicolumn{3}{c}{\textbf{RF signal}}\\
        \hline\hline
        $\omega_x/2\pi$&$3.5$~GHz&Signal carrier frequency\\
        \hline
        $B_\text{RF}$&$25$~MHz&Signal bandwidth\\
        \hline
        $T_r$&$300$~K& Reservoir temperature\\
        \hline\hline
        
        \noalign{\vskip 1em}
        \multicolumn{3}{c}{\textbf{MiRP}}\\
        \hline\hline
        $\Gamma$&\;$1$~GHz\;&MiR's total and external\\
        $\Gamma^{(\text{e})}$&\;$0.9$~GHz\;&coupling linewidth\\
        \hline
        $\Delta t$&\; 20~ns \;& Pump's time-bin spacing, \\
        $\tau$\; &\; 10~ns \;&time-bin pulse width,\\
        $\omega_w/2\pi$\; &\; 193~THz \;&carrier frequency, and its\\
        $P_{W}$&\;$1$~mW&averaged power\\
        \hline
        $C_\text{RF}$ \;& $0.7$~pF~\cite{aydin2007capacitor} &RF resonator (LC circuit) \\
        $L_\text{RF}$ \;& $0.1$~\textmu H~\cite{gely2019observation} &capacitance, inductance,\\
        $R_{\text{RF}}$ \;& $50$~$\Omega$~\cite{bagci2014optical} &radiation resistance, and its\\
        $\gamma/2\pi$&$0.5$~GHz&corresponding loss rate\\
    \hline\hline
    \end{tabular}
    \caption{Parameters of the input RF field and MiRP.}
    \label{tab:parameters}
\end{table}

\section{Detailed layout of MiRP unit}
\label{sec:EO_transducer}
Fig.~\ref{fig:scheme} illustrates the detailed MiRP unit layout. An incident RF signal is received and converted into a voltage that drives an LC resonator. To ensure efficient power transfer, the resonator impedance is set to $Z_{\text{RF}} = 377~\Omega$ to match the free-space impedance. With a radiation resistance of the LC circuit $R_{\text{RF}}$, the resonator operates with a bandwidth of $\gamma/2\pi = 0.5$~GHz and a quality factor of $Q_{\text{RF}} = 7.6$. The resulting capacitor voltage modulates the refractive index of a high-$\chi^{(2)}$ MiR. Within the MiR, a programmable optical pump $W(t)$ drives a TWM process that encodes the RF input $X(t)$ to the optical output $y(t)$. This optical signal is measured via balanced homodyne detection and digitized for inference by the digital back end.

\section{Analytical derivations in MiRP unit}
\label{sec:analytical}
In this section, we analytically derive the expression of output field of MiRP as well as its detection SNR.

\subsection{Field dynamics inside the MiR}
\label{sec:FD}
Initially, the input RF field and optical pump are injected into the MiR, where their evolution, governed by the Hamiltonians $\hat{\mathcal{H}}(t)$ and $\hat{\mathcal{E}}(t)$ in Eqs.~\eqref{eq:Hamiltonian_EO} and~\eqref{eq:Hamiltonian_encoding}. Assuming $\Gamma\gg |\dot{X}(t)/X(t)|,|\dot{W}(t)/W(t)|$ and $\gamma\gg\big|\dot{X}(t)/X(t)\big|$, we obtain the quasi-static Heisenberg–Langevin equation:
\begin{equation}
\begin{aligned}
    \frac{d\,\hat{O}(t)}{d\,t}&=\frac{i}{\hbar}\left[\hat{\mathcal{E}}(t)+\hat{\mathcal{H}}(t),\hat{O}(t)\right]-\frac{\Upsilon}{2}\,\hat{O}(t)+\hat{\mathcal{F}},
\end{aligned}
    \label{eq:master}
\end{equation}
to obtain the equations of motion for the fields $\hat{O}(t)\in\left\{\hat{a}(t),\hat{x}(t)\right\}$, and $\Upsilon\in\left\{\Gamma,\gamma\right\}$ represents the corresponding cavity decay rate. The total optical loss of the MiR is given by $\Gamma = \Gamma^{(\text{i})} + \Gamma^{(\text{e})}$, where $\Gamma^{(\text{i})}$ and $\Gamma^{(\text{e})}$ are the intrinsic and external coupling rates, respectively. Here, $$\hat{\mathcal{F}}\in\left\{\sqrt{\Gamma^{(\text{i})}}\,\hat{v}^{(\text{i})}(t)+\sqrt{\Gamma^{(\text{e})}}\,\hat{v}^{(\text{e})}(t),\sqrt{\gamma}\,\hat{\xi}(t)\right\}$$ is the Langevin noise, where $\hat{v}^{(\text{i})}(t)$, $\hat{v}^{(\text{e})}(t)$, and $\hat{\xi}(t)$ denote the vacuum noise operators associated with intrinsic, external coupling, and thermal reservoir coupling loss, respectively. These satisfy the standard correlations and commutation relations:
\begin{equation}
\begin{aligned}
    &\langle \hat{v}^{(\text{o})\dagger}(t)\,\hat{v}^{(\text{o})}(t') \rangle = 0\;\;,\;\;
\langle \hat{\xi}^\dagger(t)\,\hat{\xi}(t') \rangle = n_\text{B}\,\delta(t - t'),\\
&\big[\hat{v}^{(\text{o})}(t),\,\hat{v}^{(\text{o})\dagger}(t')\big]
= \big[\hat{\xi}(t),\,\hat{\xi}^\dagger(t')\big]
= \delta(t - t'),
\end{aligned}
\end{equation}
i.e., $\text{o}\in\{\text{i},\text{e}\}$, where $n_\text{B} = 1/\left(\exp{\left(\hbar\,\omega_x/k_\text{B}T_r\right)}-1\right)$ is the mean photon number of thermal reservoir at $\omega_x$, $T_r$ is the reservoir temperature, and $k_\text{B}$ is Boltzmann constant.

We obtain the equations of motion for $\hat{a}(t)$ and $\hat{x}(t)$ from Eq.~\eqref{eq:master}:
\begin{equation}
    \begin{aligned}
        \frac{d\,\hat{a}(t)}{d\,t}&=\frac{g}{2}\,W(t)\,\hat{x}(t)-\frac{\Gamma}{2}\,\hat{a}(t)+\sum_{\text{o}=\text{i},\text{e}}\sqrt{\Gamma^{(\text{o})}}\,\hat{v}^{(\text{o})}(t),\\
        \frac{d\,\hat{x}(t)}{d\,t}&=\frac{g}{2}\,W(t)\,\hat{a}(t)+\frac{\mu}{2}\,X(t)-\frac{\gamma}{2}\,\hat{x}(t)+\sqrt{\gamma}\,\hat{\xi}(t).
    \end{aligned}
    \label{eq:equation_of_motion}
\end{equation}
Applying adiabatic elimination, $d\,\hat{O}(t)/dt \approx 0$, we obtain the solutions as
\begin{equation}
\begin{aligned}
    \hat{a}(t)&\approx\frac{2}{\Gamma}\left(\frac{g}{2}\,W(t)\,\hat{x}(t)+\sum_{\text{0}=\text{i},\text{e}}\sqrt{\Gamma^{(\text{o})}}\,\hat{v}^{(\text{o})}(t)\right)\\
    \hat{x}(t)&\approx\frac{1}{\gamma}\,\left(g\,W(t)\,\hat{a}(t)+\mu\, X(t)+2\sqrt{\gamma}\,\hat{\xi}(t)\right)\\
    &\stackrel{(a)}{\approx}\frac{1}{\gamma}\,\left(\mu\, X(t)+2\sqrt{\gamma}\,\hat{\xi}(t)\right).
\end{aligned}
\label{eq:analytical}
\end{equation}
Approximation (a) assumes
$$g\,\left|W(t)\,\langle\hat{a}(t)\rangle\big|\ll \mu\,\big|X(t)\right|$$
(see verification in Eq.~\eqref{eq:Aa_verified} in Appendix~\ref{sec:parameter_est}).

\subsection{Output optical field in the bus waveguide}
The intracavity field $\hat{a}(t)$, once generated within the MiR, couples out into the bus-waveguide mode $\hat{b}(t)$~\cite{Wu20},
\begin{equation}
    \hat{b}(t)=\hat{v}^{(\text{e})}(t)-\sqrt{\Gamma^{(\text{e})}}\,\hat{a}(t),
    \label{eq:input_output_rel}
\end{equation}
where $\hat{v}^{(\text{e})}(t)$ denotes the vacuum input field at the external coupling port. Following the quadrature convention
\begin{equation}
    \begin{aligned}
        \hat{q}(t)&\equiv\hat{b}(t)+\hat{b}^\dagger(t)\quad\text{and}\quad\hat{p}(t)\equiv\left(\hat{b}(t)-\hat{b}^\dagger(t)\right)/i,
    \end{aligned}
\end{equation}
we obtain the output field mean quadratures:
\begin{equation}
            \langle\hat{q}(t)\rangle\approx\left(\frac{2\,\mu\,g}{\Gamma\,\gamma}\sqrt{\Gamma^{(\text{e})}}\right)\,W(t)\,X(t)\equiv y(t)\quad,\quad
            \langle\hat{p}(t)\rangle\approx0,
\label{eq:a_sol}
\end{equation}
as well as their variances: 
\begin{equation}
    \begin{aligned}
    \Delta p^2=\Delta q^2&=\int^{\infty}_{-\infty}\int^{\infty}_{-\infty}\,dt\,dt'\,f(t)\,f(t')\,\text{Cov}\left(\hat{q}(t),\hat{q}(t')\right)\\
    &=1+\frac{4\,\Gamma^{(\text{e})}}{\gamma}\,\left(\frac{g^2}{\Gamma^2}\,\left|W(t)\right|^2\right)\,\left(2\,n_\text{B}+1\right)\stackrel{(b)}{\approx} 1,\\
    \end{aligned}
\label{eq:qvar}
\end{equation}
suggesting the vacuum fluctuation variance, where $\text{Cov}(\hat{A},\hat{B})=\langle\hat{A}\hat{B}\rangle-\langle\hat{A}\rangle\langle\hat{B}\rangle$ denotes the covariance between operators $\hat{A}$ and $\hat{B}$, and $f(t)$ is the normalized temporal-mode function (i.e., $\int_{-\infty}^{\infty} |f(t)|^2 \,dt = 1$). Approximation (b) assumes $$\frac{4\,g^2\,\Gamma^{(\text{e})}}{\Gamma^2\,\gamma}\,|W(t)|^2\,(2\,n_\text{B}+1)\ll1,$$ verified in Eq.~\eqref{eq:AD_verified} in Appendix~\ref{sec:parameter_est}.

\section{Detection SNRs in MiRP and DRF sensing}
\label{sec:Transduction_eff}
\subsection{DRF sensor and $\alpha_\text{D}$}
A fully integrated DRF sensor based on a superheterodyne frequency-translating receiver remains the workhorse of modern radios, including cellular/Wi-Fi, GNSS, and radar/instrumentation systems. By translating the desired RF channel to a fixed intermediate frequency (IF), the superheterodyne architecture provides stable gain, well-defined channel selectivity, and robust blocker tolerance. This frequency translation also relaxes the dynamic-range requirements of the baseband circuitry and the analog-to-digital converter (ADC), enabling reliable performance across a wide range of operating conditions. For noise-floor benchmarking across published integrated receivers (summarized in Tab.~\ref{tab:superhet_sota}), we model the analog front end using its total input-referred noise figure ($\text{NF}$), and effective detection bandwidth $B$.

Given the noise figure NF (in dB) reported for each receiver and the reservoir temperature $T_r$, the corresponding input-referred noise temperature $T_n$ and the integrated noise power $P_{\text{noise}}$ over bandwidth $B$ are
\begin{equation}
    T_n = T_r\,10^{\text{NF}/10}\quad,\quad P_\text{noise} = k_\text{B}\,T_n\,B.
\label{eq:Pnoise_def}
\end{equation}
The resulting detection SNR is therefore
\begin{equation}
\text{SNR}_\text{D}= \alpha_\text{D}\,P_{\text{RF}},
\label{eq:SNR_def}
\end{equation}
where
\begin{equation}
    \alpha_\text{D}\equiv \frac{1}{P_\text{noise}} = \frac{1}{k_\text{B} T_r\,10^{\text{NF}/10}\,B}
    \label{eq:alpha}
\end{equation}
denotes the power-to-SNR coefficient.

Tab.~\ref{tab:superhet_sota} compiles state-of-the-art parameters reported for integrated superheterodyne receivers. For each design, the reported $\text{NF}$ and detection bandwidth $B$ are extracted, and the corresponding power-to-SNR coefficient $\alpha_\text{D}$ is computed using Eq.~\eqref{eq:alpha}.

\begin{table}[t]
\begin{ruledtabular}
\begin{tabular}{c c c c c}
 Ref.& NF & $B$  &  $P_\text{noise}$ & $\alpha_\text{D}$\\
&(dB)&(MHz)&(pW) &(pW$^{-1}$)\\
\hline
\cite{zijlma2024}                  & 4.7--7.0   &  20 &  0.24--0.42 & 2.41--4.09 \\
\cite{vanzanten2024isscc}          & 4.8--5.5   &  20 &  0.25--0.29 & 3.40--4.00 \\
\cite{tao2024}                     & 3.5--5.7   &  50 &  0.46--0.77 & 1.30--2.16 \\
\cite{vanzanten2025jssc}           & 5.1--9.0   &  40 & 0.54--1.32  & 0.76--1.87 \\
\cite{seo2021jssc}                 & 5.5--7.1   &  65 & 0.96--1.38  & 0.72--1.05 \\
\cite{razavi2022}                  & 2.1--4.42  & 160 &  1.07--1.83 & 0.55--0.93 \\
\cite{montazerolghaem2023isscc}    & 3.2--5.8   & 300 & 2.60--4.72  & 0.21--0.39 \\
\cite{seo2021rfic}                 & 9.7        & 170 & 6.6         & 0.15 \\
\end{tabular}
\end{ruledtabular}
\caption{State-of-the-art fully integrated frequency-translating receivers used to benchmark DRF. $\alpha_\text{D}$ is calculated via Eq.~\eqref{eq:alpha}.}
\label{tab:superhet_sota}
\end{table}

For the DRF baseline simulations in Fig.~\ref{fig:performance} in the main text, we use the highest reported detection coefficient in Tab.~\ref{tab:superhet_sota},
\begin{equation}
    \alpha_\text{D} = 4.09~\text{pW}^{-1}.
    \label{eq:alpha_D}
\end{equation}
This value represents the state-of-the-art performance limit of modern integrated superheterodyne receivers and provides a consistent electronic noise-floor benchmark for the comparisons in the main text.

\begin{table}[t]
\label{tab:summary}
\begin{ruledtabular}
\begin{tabular}{c @{\hspace{1.1em}} c @{\hspace{1.1em}} c @{\hspace{1.1em}} c @{\hspace{1.1em}} c @{\hspace{1.1em}} c}
Ref.&Platform&$\omega_x/2\pi$&$B_\text{T}$&$\beta$&$\alpha_\text{M}$\\
&&(GHz)&(MHz)&(\%)&(pW$^{-1}$)\\
\hline
\cite{brubaker2022optomechanical}&NbTiN&$7.9$&$0.002$&$47$&$110$\\
\cite{Andrews2014Conversion}&LiNbO$_3$&$7$&$0.03$&$8.6$&$22$\\
\cite{sahu2022quantum}&LiNbO$_3$&$8.8$&$10$&$8.7$&$18$\\
\cite{warner2023coherent}&LiNbO$_3$&$3.5$&$30$&$1.18$&$6.2$\\
\cite{xu2021bidirectional}&LiNbO$_3$&$7.7$&$9.1$&$1.02$&$2.4$\\
\cite{han2018coherent}&$^{87}$Rb&$84$&$4$&$0.3$&$0.066$\\
\cite{hease2020bidirectional}&LiNbO$_3$&$9$&$10.7$&$0.03$&$0.062$\\
\end{tabular}
\end{ruledtabular}
\caption{State-of-the-art parameters of RF-to-optics transducers. $B_\text{T}$: RF-to-optics transduction bandwidth, i.e., $\gamma/2\pi=0.5$~GHz (Tab.~\ref{tab:parameters}). $\alpha_\text{M}$ is calculated via Eq.~\eqref{eq:alpha_M}.}
\label{tab:efficiency}
\end{table}

\subsection{MiRP and $\alpha_\text{M}$}
\label{sec:MiRP_alpha}
In MiRP (or RFP), balanced homodyne readout employs a strong local oscillator (LO) to measure the $q$-quadrature of the transduced optical field $\hat{q}(t)$; the operator for the output difference electron number in the mode duration $\tau$ is written as:
\begin{equation}
\hat{n}(t) = \sqrt{n_L}\,\hat{q}(t) + \varepsilon_\text{HR}(t),
\label{eq:observable_HR}
\end{equation}
where $n_L$ represents the mean photon number of the LO and $\varepsilon_{\text{HR}}(t)$ denotes the noise per temporal mode. The time-averaged detection SNR over period $T$ is therefore calculated as:
\begin{equation}
\begin{aligned}
    \text{SNR}_\text{M}&=\frac{\frac{1}{T}\int_0^T\langle\hat{n}(t)\rangle^2\,dt}{\Delta n^2}\stackrel{(c)}{\approx}\alpha_\text{M}\,P_\text{RF},
\end{aligned}
\label{eq:SNR_LM}
\end{equation}
where
\begin{equation}
\alpha_\text{M}\equiv4\,\beta/\hbar\omega_x\gamma\quad\text{and}\quad\beta=\Gamma^{(\text{e})}\mu^2g^2\left|W_{\text{rms}}\right|^2/\gamma^2\Gamma^2,
    \label{eq:alpha_M}
\end{equation}
i.e., $W_{\text{rms}}\equiv \sqrt{\frac{1}{T}\int_{0}^T\,|W(t)|^2\,dt}$ is the root-mean-square of the pump. Approximation (c) assumes $\Delta\varepsilon^2_\text{HR}/n_L \ll 1$, numerically validated in Eq.~\eqref{eq:HM_shotnoise} in Appendix~\ref{sec:parameter_est}. In this framework, $\alpha_\text{M}$ and $\beta$ represent the power-to-SNR coefficient and the RF-to-optics transduction efficiency, respectively. 

With the phase-sensitive amplification (PSA) gain $G$, the power-to-SNR coefficient becomes:
\begin{equation}\alpha_\text{M} \rightarrow \alpha_\text{M}^{(G)} = \left( \sqrt{G} + \sqrt{G-1} \right)^2 \alpha_\text{M}.
\label{eq:Gain_effect}
\end{equation}
This parametric amplification allows the system to surpass classical sensitivity limits, boosting the SNR without increasing the physical intensity of the signal or the pump.

\begin{figure}[h]
    \centering
{\centering\includegraphics[width=0.95\linewidth]{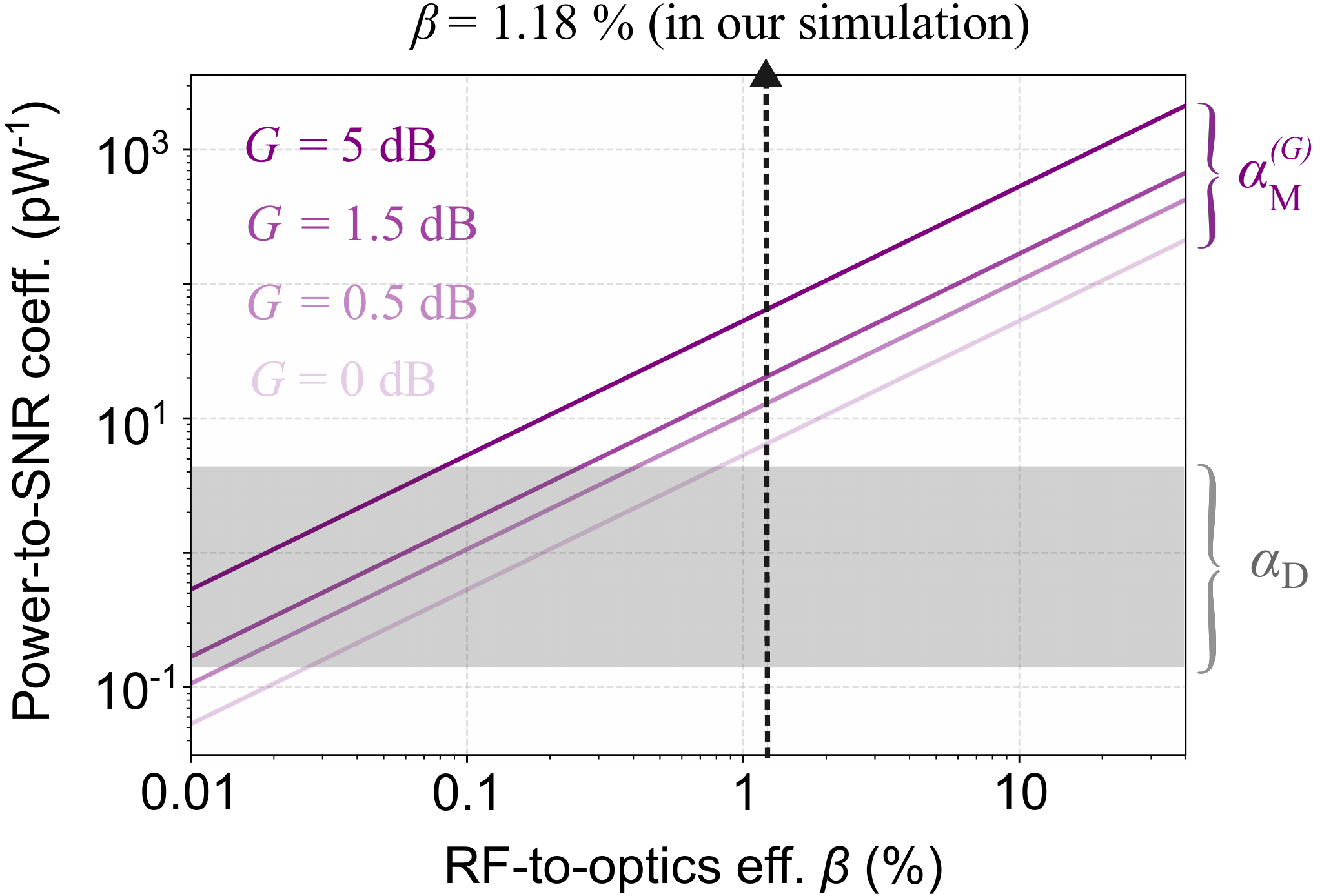}}
	\caption{Power-to-SNR coefficients. $\alpha_\text{M}^{(G)}$ versus RF-to-optics transduction efficiency $\beta$ for PSA gains $G = 0, 0.5, 1.5, 5$ dB; the state-of-the-art DRF's $\alpha_\text{D}$ range (summarized in Tab.~\ref{tab:superhet_sota}) is shown for reference.}
	\label{fig:DRF_RFP}
\end{figure}

The sensing performance of MiRP is fundamentally influenced by the RF-to-optics transduction efficiency $\beta$, as defined in Eq.~\eqref{eq:alpha_M}. Since MiRP and DRF operate on different physical platforms, their effective SNR coefficients depend on distinct device parameters. Tab.~\ref{tab:efficiency} summarizes representative state-of-the-art $\beta$s reported in recent literature, alongside their corresponding power-to-SNR coefficients $\alpha_\text{M}$.

As illustrated in Fig.~\ref{fig:DRF_RFP}, the power-to-SNR coefficient $\alpha_{\text{M}}^{(G)}$ increases with both the transduction efficiency $\beta$ and the PSA gain $G$. When either parameter is sufficiently large, the resulting $\alpha_{\text{M}}^{(G)}$ outperforms the DRF benchmark $\alpha_{\text{D}}$, whose reported range $0.15~\text{pW}^{-1} \le \alpha_{\text{D}} \le 4.09~\text{pW}^{-1}$ is summarized in Tab.~\ref{tab:superhet_sota}. This hardware-level SNR advantage explains why RFP in Fig.~\ref{fig:performance} outperforms DRF. MiRP benefits from the same hardware gain and, in addition, further improves performance through its learned analog preprocessing.

\section{Validity of assumptions in MiRP}
\label{sec:parameter_est}
In this section, we numerically validate the assumptions used in Eqs.~\eqref{eq:analytical},~\eqref{eq:qvar},~\eqref{eq:SNR_LM}.

\subsubsection{Coherent optical pump}
The optical pump $W(t)$ is structured in time bins, each has width $\tau=1$~ns and average power $P_W = 1$~mW. The corresponding mean photon number per bin is
\begin{equation}
\frac{P_W\,\tau}{\hbar\,\omega_w} = \left|W_{\text{rms}}\right|^2 \sim 4\times10^8,
\end{equation}
which satisfies Approximation (a) in Eq.~\eqref{eq:analytical}:
\begin{equation}
\begin{aligned}
    \frac{g\,\left|W(t)\,\langle\hat{a}(t)\rangle\right|}{\mu \,\left|X(t)\right|}&\approx\frac{g\,\left|W(t)\right|^2}{\gamma\,\Gamma}\approx\frac{g\,\left|W_{\text{rms}}\right|^2}{\gamma\,\Gamma}\sim 5\times10^{-8},
\end{aligned}
\label{eq:Aa_verified}
\end{equation}
and (b) in Eq.~\eqref{eq:qvar}:
\begin{equation}
\begin{aligned}
    \frac{4\,g^2\,\Gamma^{(\text{e})}}{\Gamma^2\,\gamma}\,|W(t)|^2\,(2\,n_\text{B}+1)&\approx \frac{4\,g^2\,\Gamma^{(\text{e})}}{\Gamma^2\,\gamma}\,\left|W_{\text{rms}}\right|^2\,(2n_\text{B}+1)\\
    &\sim 10^{-1},
\end{aligned}
\label{eq:AD_verified}
\end{equation}
i.e., $g/2\pi=100$~Hz~\cite{Holzgrafe:20} and $n_\text{B}\sim1.7\times10^3$.

\subsubsection{Local oscillator in homodyne}
In optical homodyne detection, we set the detector’s electronic noise and LO powers to $P_e=-70$ dBm and $P_\text{LO}=25$ mW~\cite{Vahlbruch16}. Under these conditions, the LO shot noise dominates the detector noise, with power
\[
P_\text{shot} = \frac{2\,q_e\,G^2\,\mathcal{R}\,P_\text{LO} \, \Delta f}{Z_\text{HR}} = -42~\text{dBm},
\]
where $G = 4$ k$\Omega$ is the transimpedance gain, $\mathcal{R} = 1.24$ A/W is the photodiode responsivity, $q_e$ is the electron charge, $Z_\text{HR} = 50~\Omega$ is the detector impedance, and $\Delta f = 20$ MHz is the HR bandwidth. The ratio of electronic noise to shot-noise power is therefore
\begin{equation}
\frac{P_e}{P_\text{shot}} = \frac{\Delta\varepsilon^2_\text{HR}}{n_L} = 1.6\times 10^{-3},
\label{eq:HM_shotnoise}
\end{equation}
which justifies Approximation (c) in Eq.~\eqref{eq:SNR_LM}.

\section{Sensing tasks and model configurations}
\label{sec:tasks}
The sensing tasks discussed in the main text include MNIST~\cite{lecun2002gradient}, human activity recognition~\cite{anguita2013har}, and MNIST-1D~\cite{greydanus2024scaling}. These tasks are described in the following subsections, with each dataset characterized by the input size $M$, the number of input channels $J$, and the number of classes $\mathcal{C}$.

All sensing pipelines employ a digital back end consisting of two convolutional (CONV) layers followed by three fully connected (FC) layers, with layer dimensions treated as task-dependent hyperparameters that are independently optimized and differ between the MiRP and conventional sensing pipelines.

\subsection{MNIST}
A canonical handwritten digit dataset~\cite{lecun2002gradient}, comprising $10$ classes ($0$--$9$, with $\mathcal{C}=10$), is used for evaluation. Each grayscale image has an original size of $28\times28$ and is flattened into a one-dimensional array with $(J, M) = (1, 784)$ elements to match the one-dimensional input format of our sensing system. 

For this task, MiRP employs CONV layers $(1,128)$ and $(1,128)$, followed by FC layers $(512)$, $(256)$ and $(10)$, whereas the conventional pipeline uses CONV layers $(1,256)$ and $(1,256)$, followed by FC layers $(512)$, $(256)$, and $(10)$.


\subsection{Human posture detection}
A physical sensor dataset~\cite{anguita2013har}, comprising experimental time-series data from six human activities (walking, walking upstairs, walking downstairs, sitting, standing, and laying down), is used for evaluation ($\mathcal{C}=6$). Each sample has a size of $(J, M) = (1, 561)$ and is collected using a body-worn accelerometer and gyroscope to measure acceleration and angular velocity along three spatial axes. 

For this task, MiRP uses CONV layers $(1,128)$ and $(1,128)$, followed by FC layers $(256)$, $(128)$, and $(6)$, while the conventional pipeline uses CONV layers $(1,256)$, $(1,256)$, followed by FC layers $(1024)$, $(512)$, and $(6)$.

\subsection{MNIST-1D}
The one-dimensional analog of the MNIST database (MNIST-1D)~\cite{greydanus2024scaling} comprises synthetic time-series waveforms representing handwritten digits ($0$--$9$, with $\mathcal{C}=10$). Each sample is a single-channel signal of size $(J, M) = (1, 40)$, generated by embedding a digit template into structured noise and applying random transformations such as scaling, translation, and temporal distortion to emulate variability in sequential measurements. 

For this task, MiRP uses CONV layers $(1,128)$ and $(1,128)$ followed by FC layers $(512)$, $(256)$, and $(10)$, respectively, whereas the conventional pipeline uses CONV layers $(1,256)$ and $(1,256)$ followed by FC layers $(1024)$, $(512)$, and $(10)$, respectively.

\bibliographystyle{apsrev4-2}
\bibliography{References}
\end{document}